\documentclass[12pt]{article}
\usepackage{amssymb,graphicx}
\usepackage{epstopdf}
\usepackage{amsmath,amsfonts}
\usepackage{epsfig}
\RequirePackage{ifpdf}
\ifpdf

\def\ext{pdf}
\usepackage[pdftex]{hyperref}
\else

\def\ext{eps}
\usepackage[hypertex]{hyperref}
\fi

\def\e{{\rm e}}

\def\d{\partial}
\def\l{\left(}
\def\r{\right)}

\newcommand{\be}{\begin{equation}}
\newcommand{\ee}{\end{equation}}
\newcommand{\bea}{\begin{eqnarray}}
\newcommand{\eea}{\end{eqnarray}}
\newcommand{\bg}{\begin{gather}}
\newcommand{\eg}{\end{gather}}
\newcommand{\bseq}{\begin{subequations}}
\newcommand{\eseq}{\end{subequations}}

\begin{document}
\begin{flushright}
INR-TH-2011-10
\end{flushright}
\vspace{10pt}
\begin{center}
  {\LARGE \bf Non-Gaussianity of scalar perturbations \\[0.3cm]
generated by conformal mechanisms} \\
\vspace{20pt}
M.~Libanov, S.~Mironov, V.~Rubakov\\
\vspace{20pt}
\textit{
Institute for Nuclear Research of
         the Russian Academy of Sciences,\\  60th October Anniversary
  Prospect, 7a, 117312 Moscow, Russia;}\\
\vspace{10pt}
\textit{
Physics Department, Moscow State University\\ Vorobjevy Gory,
119991, Moscow, Russia
}

\end{center}

    \vspace{5pt}

\begin{abstract}
We consider theories which explain the flatness of the power spectrum of
scalar perturbations in the Universe by conformal invariance, such as
conformal rolling model and Galilean Genesis. We show that to the leading
{\it non-linear} order, perturbations in all models from this class behave
in one and the same way, at least if the energy density of the relevant
fields is small compared to the total energy density (spectator
approximation). We then turn to the intrinsic non-Gaussianities in these
models (as opposed to non-Gaussianities that may be generated during
subsequent evolution). The intrinsic bispectrum vanishes, so we perform
the complete calculation of the trispectrum and compare it with the
trispecta of local forms in various limits. The most peculiar feature of
our trispectrum is a (fairly mild) singularity in the limit where two
momenta are equal in absolute value and opposite in direction (folded
limit). Generically, the intrinsic non-Gaussianity can be of detectable
size.
\end{abstract}

\section{Introduction}
\label{sec:Intro}

Scalar perturbations in the Universe may well originate from
inflation~\cite{infl-perturbations}. In that case, approximate flatness of
their power spectrum is due to the approximate de~Sitter symmetry of
inflating background. As pointed out in Ref.~\cite{Antoniadis:1996dj}, a
possible alternative to the de~Sitter symmetry in this context is
conformal invariance. Concrete models employing conformal invariance,
instead of the de~Sitter symmetry, for explaining the flat scalar
spectrum include conformal
rolling~\cite{vrscalinv} and Galilean Genesis~\cite{Creminelli:2010ba}.
Despite substantially different motivations and dynamical features, the
latter models end up in one and the same mechanism that generates scalar
perturbations. From this prospective, the theory
boils down to a model with two scalar fields $\rho$ and $\Theta$ and the
Lagrangian
\be
L = L_\rho + \frac{1}{2} \rho^2 \l \d_\mu \Theta \r^2 \; ,
\label{may17-2}
\ee
where $L_\rho$ governs  the dynamics of $\rho$.
Under scaling transformations,
the field $\rho$ scales as $\rho(x) \to \lambda \rho(\lambda x)$; it
provides for a non-trivial background $\rho_c$.
The field $\Theta$ scales as $\Theta (x) \to \Theta (\lambda x)$;
  the perturbations of the field
$\Theta$ serve as predecessors of the adiabatic  perturbations. One makes
sure that at the time the perturbations of $\Theta$ are generated,
gravitational effects on the dynamics of the two fields are irrelevant
(this is achieved in different ways in Ref.~\cite{vrscalinv}
and Ref.~\cite{Creminelli:2010ba})
and
assumes that the background field $\rho_c$ is spatially homogeneous. Then
conformal invariance of the field equation for $\rho$ implies that
\be
\rho_c (x_0) = - \frac{1}{x_0} \; , \;\;\;\;\;\; x_0 < 0 \; ,
\label{may15-1}
\ee
where $x_0=\eta $ is conformal time in the conformal rolling model and
$x_0=t$ is cosmic time in the Galilean Genesis scenario. In such a
background, the field $\Theta$ behaves exactly in the same way as massless
scalar field (minimally coupled to gravity) in the de~Sitter space-time.
The modes of its perturbations  about a certain background value
$\bar{\Theta}$,
\[
\theta (x) = \Theta (x) - \bar{\Theta} \; ,
\]
start off in the WKB regime and freeze out when $k| x_0| \sim
1$, where $k$ is (conformal) momentum. Assuming that the field
$\theta (x)$
 is
originally in its vacuum state,
one obtains, at the linearized level, the
flat power spectrum of the Gaussian field $\theta ({\bf k})$ at late
times\footnote{Deviation from exact conformal invariance naturally gives
rise to the tilt in the power spectrum, see, e.g., Ref.~\cite{Osipov}.},
when $k|x_0| \ll 1$. The conformal stage ends up at some point, and the
$\theta$-perturbations are reprocessed into adiabatic ones at much later
epoch via, e.g., curvaton~\cite{Linde:1996gt} or modulated
decay~\cite{Dvali:2003em} mechanism.

A common feature of the conformal models of
Refs.~\cite{vrscalinv,Creminelli:2010ba} is the existence of the
perturbations of the field $\rho$ about the background \eqref{may15-1}.
The Lagrangian $L_\rho$ contains a small parameter, call it $h$, then
$\delta \rho \propto h$ ($h^2$ is quartic self-coupling in the conformal
rolling model; we relate the parameter $h$ to the parameters of
the galileon Lagrangian in the Galilean Genesis scenario in
Section~\ref{sec:galileon}, see Eq.~\eqref{may26-1}). As discussed in
Refs.~\cite{vrscalinv,Creminelli:2010ba}, the perturbations $\delta \rho$
have rather peculiar properties. Nevertheless, as we review in
Section~\ref{sec:galileon}, these perturbations are exactly the same
(modulo overall amplitude) in the two, apparently very different models.
Furthermore, in Section \ref{sec:general} we give a general argument
showing that the properties of the linearized perturbations $\delta \rho$
are common to a large class of models employing conformal invariance.

There is a subtlety here. Strictly speaking, our observation is
valid in a theory without gravity or in the case when the energy density
of the field $\rho $ is small compared to the total energy density in the
Universe (spectator approximation). Though is is likely that our results
are valid in much more general setting, the corresponding analysis is yet
to be done. We proceed in this paper by neglecting gravity altogether.

So, it is of interest to study the effects of the perturbations $\delta
\rho$ on the field $\theta$ and, in the end, on the adiabatic
perturbations. There are at least two of these effects, namely, the
statistical anisotropy and non-Gaussianity. Of course, the non-Gaussianity
may well be generated also at the time the $\theta$-perturbations are
converted into the adiabatic ones. In this respect the conformal models
are nothing special as compared to inflationary theories equipped with the
curvaton or modulated decay mechanism; for this reason we are going to
disregard this conversion-related non-Gaussianity. What we are interested
in is the intrinsic non-Gaussianity, which is due to the interaction of
the field $\theta$ with perturbations $\delta \rho$.

The resulting phenomenology depends strongly on what happens to the field
$\theta$ after the conformal stage \eqref{may15-1} ends up. One option is
that the field $\theta({\bf k})$ starts evolving again, and its evolution
proceeds until it becomes superhorizon in the conventional sense. This
option is fairly natural in the conformal rolling model of
Ref.~\cite{vrscalinv} and more contrived in the Galilean
Genesis\footnote{In the context of the Galilean Genesis, an intermediate
stage of the evolution of $\theta$ may occur provided that the effective
scale factor $\rho(\pi)$ is a non-trivial function of the galileon field
$\pi$, such that $\rho \propto \e^\pi$ at $\pi$ smaller than some value
$\pi_0$ and $\rho = \mbox{const}$ at $\pi>\pi_0$. Then the field $\theta$
feels the background \eqref{may15-1} at early times, when $\pi<\pi_0$, and
temporarily gets frozen out when $k|x_0| \sim 1$, as discussed in the
text. If the opposite regime $\pi>\pi_0$ sets in later, but still at some
sufficiently early time when the space-time is nearly Minkowskian,
the field $\theta ({\bf k}) $ indeed starts to oscillate again at that
time. These oscillations terminate when the Hubble parameter becomes large
enough and the field $\theta ({\bf k})$ exits the horizon.} of
 Ref.~\cite{Creminelli:2010ba}. Both statistical anisotropy and
non-Gaussianity generated in this case are studied in
Ref.~\cite{Libanov:2011hh}.

Here we consider the opposite case, i.e., we assume that the field
$\theta$ does not evolve after the end of the conformal stage. This option
is particularly natural in the Galilean Genesis, but it is not contrived
in the conformal rolling scenario either. In models from this sub-class, the
adiabatic perturbations inherit the statistical properties of
$\theta$-perturbations that exist already at the late conformal stage,
when $k|x_0| \ll 1$. The statistical anisotropy in the field $\theta$ is
induced by long-ranged perturbations $\delta \rho$. This effect has been
studied in Ref.~\cite{Libanov:2010nk} with the result that
the leading statistical anisotropy in
the power
spectrum of $\theta$, and hence of adiabatic perturbation $\zeta$, has the
quadrupole form,
\[
{\cal P}_\zeta ({\bf k}) = {\cal P}_0 (k) \left(1 + c_1 \cdot h \cdot
\frac{H_0}{k} \cdot \frac{{k}_i {k}_j}{k^2} w_{ij} - c_2 \cdot h^2 \cdot
\frac{({\bf   k u})^2}{k^2}\right)\; ,
\]
where ${\cal P}_0 (k)$ is isotropic (and nearly flat) power spectrum
obtained at the linearized level, $u_i$ and $w_{ij}$ are unit 3-vector and
unit traceless 3-tensor of a general form, $H_0$ is the present value of
the Hubble parameter, the parameter $c_1$ is of order 1 and $c_2$ is
positive and
logarithmically enhanced.

The main purpose of this paper is to study the
intrinsic (as opposed to conversion-related) non-Gaussianity in this
sub-class of models.
In the absence of the cubic self-interaction of the field $\theta$, the
intrinsic bispectrum vanishes, so we have to consider the intrinsic
trispectrum. Perhaps the most striking feature of the trispectrum is the
singularity in the limit where two momenta are equal in absolute value and have
opposite directions (folded limit, in nomenclature of
Ref.~\cite{Chen:2009bc}). The singular part of the connected four-point
function has been calculated in our earlier
paper~\cite{Libanov:2010ci} with the result
\begin{align}
\langle \zeta_{\mathbf{k}_{1}}  \zeta_{\mathbf{k}_{2}}
\zeta_{\mathbf{k}_{3}} \zeta_{\mathbf{k}_{4}} \rangle = \mbox{const} &
 \cdot \delta \left(\sum \limits_{i=1}^{n}\mathbf{k}_{i}\right) \cdot
\frac{1}{k_{12}k_1^{4}k_3^{4}} \left[1-3\left( \frac{\mathbf{k_{12}
k_1}}{k_{12} k_1} \right)^{2} \right] \left[1-3\left(\frac{\mathbf{k_{12}
k_3}}{k_{12} k_3} \right)^{2} \right]
\label{may21-1}\\
\mathbf{k_{12}} &= \mathbf{k_{1}} + \mathbf{k_{2}} \to 0 \; , \nonumber
\end{align}
i.e., the trispectrum blows up as $k_{12}^{-1}$. This is in contrast to
trispectra obtained in single-field inflationary
models~\cite{Chen:2009bc,Seery:2006vu,Seery:2008ax,Arroja:2009pd,Bartolo:2010di},
and, indeed, there are general arguments~\cite{Seery:2008ax} showing that
in these models, the four-point function is finite in the limit
$k_{12} \to 0$. The singularity in the four-point function \eqref{may21-1}
is due to the enhancement of the perturbations $\delta \rho$ at low
momenta, see the discussion of the infrared properties of the conformal
models in Ref.~\cite{Libanov:2010nk}. We will see in
Section~\ref{sec:shapes} that the most relevant features of the trispectrum
are captured by its singular part.

Whether the intrinsic non-Gaussianity dominates over the conversion-related
one, and whether the former is detectable depends on both the
underlying conformal model and the mechanism that converts the
$\theta$-perturbations into the adiabatic ones. Generically,
the linear order relationship between $\zeta$ and $\theta$ is
$\zeta = r \theta/\bar{\Theta} $,
where $r \lesssim 1$ is the dilution factor. With our normalization
(\ref{may17-2}), (\ref{may15-1}), the power spectrum of $\theta
$-perturbations is $\mathcal{ P}_{\theta }=1/(4\pi ^{2})$. The background
value $\bar{\Theta }$ cannot be estimated in a model independent way. It
can be as large as $\bar{\Theta }\sim 10^{4}$, so that the correct
adiabatic amplitude is obtained at $r\sim 1$. This can be the case, e.g.,
in the Galilean Genesis model. In such a situation, the conversion-related
non-Gaussianities induced by the curvation or modulated decay mechanism
are fairly small ($f_{NL}$, $g_{NL}$ are roughly of order
1~\cite{Lyth:2002my,Zaldarriaga:2003my,Ichikawa:2008ne,Byrnes:2010em}). On
the other hand, the size of the intrinsic non-Gaussianity $t_{NL}$ (see
Eq.~(\ref{aug31-1}) for its definition) is governed by the amplitude of
the perturbations $\delta \rho$, so that
\[
t_{NL} \sim \frac{h^2}{4\pi^2 {\cal P}_\zeta} \; ,
\]
see Eq.~(\ref{aug31-2}) for numerical coefficient. There is no general
reason to expect that the parameter $h$ is particularly small, except for
the mild requirement $h \ll 1$ ensuring the self-consistency of the
conformal scenario. This shows  that the intrinsic non-Gaussianity may
well be dominant and detectable (say, $t_{NL} \gtrsim 10^6$), though it
size cannot be predicted because of our ignorance of the value of the
parameter $h$. The situation is more subtle in the conformal rolling
model; we consider this point in Section~\ref{sec:shapes} with the result
that dominant and detectable intrinsic non-Gaussianity is possible, but
not generic.

This paper is organized as follows. To set the stage, we consider in
Section~\ref{sec:galileon} the linearized perturbations $\theta$ and
$\delta \rho$ in the Galilean Genesis scenario and compare the resulting
expressions with those obtained in the conformal rolling model. This
Section contains nothing new as compared to
Refs.~\cite{vrscalinv,Creminelli:2010ba}; the main point is to show that
the perturbations $\theta$ and $\delta \rho $ are identically the same in
the two scenarios. In Section~\ref{sec:general} we give a general argument
showing that the properties of $\delta \rho$ are uniquely determined by
conformal invariance (modulo the overall constant amplitude). Hence, the
peculiarities of the adiabatic perturbations that we study in this paper,
as well as the results of Refs.~\cite{Libanov:2010nk,Libanov:2010ci}, are
common to the whole class of conformal mechanisms. We turn to the
non-Gaussianity in Section~\ref{sec:main}, where we perform the complete
calculation of the intrinsic trispectrum. We confirm our earlier
result~\eqref{may21-1} concerning the singular part of the trispectrum. We
then consider various limits, following the nomenclature of
Refs.~\cite{Chen:2009bc,Bartolo:2010di}, and compare them with local
models~\cite{Okamoto:2002ik,Kogo:2006kh}. We conclude in
Section~\ref{sec:concl}. Some details of our calculations are collected in
Appendix.

\section{Perturbations in conformal scenarios}
\label{Section/Pg5/1:paper/Perturbations in conformal scenarios}

\subsection{Galilean Genesis vs. conformal rolling}
\label{sec:galileon}

The galileon model has been introduced in Ref.~\cite{Nicolis:2008in}. The
rolling galileon serves as the field $\rho$ entering \eqref{may17-2} and
\eqref{may15-1}. In the Minkowski space-time, the Lagrangian of the
simplest conformally-invariant version~\cite{Creminelli:2010ba} of the model
 is (mostly negative signature)
\be
L_\pi = - f^2 \e^{2\pi} \d_\mu \pi \d^\mu \pi + \frac{f^3}{\Lambda^3}
\d_\mu \pi \d^\mu \pi \Box \pi + \frac{f^3}{2\Lambda^3} (\d_\mu \pi \d^\mu
\pi)^2 \; ,
\label{may24-3}
\ee
where $\Box = \d_\mu \d^\mu$. The field equation in Minkowski space-time
admits the
homogeneous solution $\pi_c$ such that
\be
\e^{\pi_c} = - \frac{1}{H_* x_0} \; ,
\label{may17-5}
\ee
where $x_0 = t$ and
\[
H_*^2 = \frac{2\Lambda^3}{3 f} \; .
\]
Hence, one defines
\be
\rho = H_* \e^{\pi}
\label{may17-1}
\ee
so that the background solution is given by \eqref{may15-1}.

The quadratic action for perturbations about this solution
is
\[
S_{\delta \pi} = \frac{f^2}{H_*^2} \int~d^4x~ \left\{ \frac{1}{x_0^2}
[(\d_0 \ {\delta \pi})^2 - (\d_i \ \delta \pi)^2] + \frac{4}{x_0^4} \delta
\pi^2 \right\}\;.
\]
It is convenient to introduce the variable $\delta \rho = - x_0^{-1}
\delta \pi$. It follows from \eqref{may17-1} that $\delta \rho$ is the
perturbation of the field $\rho$ about the background $\rho_c =
-x_0^{-1}$. Its action reads
\[
S_{\delta \rho} = \frac{2f^2}{H_*^2} \int~d^4x~\frac{1}{2} \left[
(\d_0\ {\delta \rho})^2 - (\d_i \ \delta \rho)^2 + \frac{6}{x_0^2} (\delta
\rho)^2\right]\;.
\]
The field equation has a simple form,
\be
\d_0^2{\delta \rho} - \Delta \delta \rho - \frac{6}{x_0^2} \delta \rho
= 0 \; .
\label{may18-21}
\ee
Its properly normalized solution for given momentum ${\bf k}$ is
\be
\delta \rho = h\cdot \mbox{e}^{i{\bf kx}} \cdot \frac{i}{4\pi}
\sqrt{\frac{-t}{2}} H_{5/2}^{(1)} \left(- k x_0 \right) \cdot \hat{B}_{\bf
k} + h.c.\; ,
\label{apr26-3}
 \ee
 where $ \hat{B}^\dagger_{\bf k}$  and $\hat{B}_{\bf k}$ are creation and
 annihilation operators obeying the standard commutational relation
 $[\hat{B}_{\bf k}, \hat{B}^\dagger_{\bf q}] = \delta({\bf k} - {\bf q})$,
 and
\be
h = \frac{H_*}{\sqrt{2}f} \; .
\label{may26-1}
\ee
The theory is weakly coupled in the relevant range of momenta provided
that $h \ll 1$.

The mode \eqref{apr26-3} oscillates at early times, when $k|x_0| \gg 1$,
while at late times it behaves as follows,
 \[
\delta \rho = h \cdot\mbox{e}^{i{\bf kx}} \cdot \frac{3}{4\pi^{3/2}}
\frac{1}{k^{5/2} x_0^2} \cdot \hat{B}_{\bf k} + h.c.\; .
\]
This behavior can  be interpreted as local time
shift~\cite{vrscalinv,Creminelli:2010ba}. Indeed, for time-shifted
galileon solution \eqref{may17-5} we have
\[
\e^{\pi_c + \delta \pi} = \e^{\pi_c} (1+ \delta \pi)= - \frac{1}{H_*
(x_0+\delta x_0)} = - \frac{1}{H_* x_0} + \frac{\delta x_0}{H_* x_0^2} \;
.
\]
Hence,
\[
\delta x_0 = - x_0 \delta \pi =  x_0^2 \delta \rho  \;.
\]
Thus, $\delta x_0 ({\bf x})$ is independent of time at late times, and
its power spectrum is red,
\be
{\cal P}_{\delta x_0} = \frac{9h^2}{4\pi^2} \frac{1}{k^2} \; .
\label{may18-20}
\ee

In the Galilean Genesis scenario, the field $\Theta$ is introduced as an
additional field precisely for the purpose of generating the scalar
 perturbations. By conformal invariance, its quadratic Lagrangian has the
form of the second term in \eqref{may17-2}. In the background $\rho_c =
-x_0^{-1}$, its modes are
\be
\theta = \mbox{e}^{i{\bf kx}} \cdot \frac{1}{4\sqrt{2} \pi}
(-x_0)^{3/2} H^{(1)}_{3/2} (-kx_0) \cdot \hat{A}_{\bf k} + h.c. \; ,
\label{may18-1}
\ee
where $ \hat{A}^\dagger_{\bf k}$  and $\hat{A}_{\bf k}$ is another set
of creation and annihilation operators. At early times, the field $\theta$
is in the WKB regime, while for $k|x_0| \ll 1$ the mode stays constant in
time. The resulting late-time power spectrum is flat,
\be
{\cal P}_\theta = \frac{1}{4\pi^2} \; .
\label{may18-50}
\ee
This result is valid at the linearized level. The lowest order
interaction between $\theta$ and $\delta \rho$ is described by the
interaction Hamiltonian, whose density is (in the interaction picture)
\begin{equation}
\mathcal{ H}_{I}=- L_{\mathrm{int}}= - \rho _{c}\ \delta \rho \
(\d_\mu \theta)^2 \; .
\label{Eq/Pg3/3:in-in}
\end{equation}
It is this interaction that is responsible for the intrinsic
non-Gaussianity which we study in Section~\ref{sec:main}.

In the above discussion we neglected gravity effects. This is legitimate
in the Galilean Genesis scenario at the early stage of Genesis, when the
energy density of the galileon field is small, while the energy density of
other fields is assumed to vanish. The latter assumption is relaxed in the
conformal rolling scenario whose main ingredient is a complex scalar field
$\phi$ {\it conformally} coupled to gravity. Unlike in the galileon case,
the scalar potential does not vanish; it is assumed to be negative and
is quartic by conformal symmetry, $V(\phi) = - h^2 |\phi|^4$. One makes
use of the parametrization
\be
\phi = \frac{\rho}{h} \mbox{exp} \l i \frac{h \Theta}{\sqrt{2}} \r \; .
\label{sep1-2}
\ee
Then the background field $\rho_c$ rolls down the potential according
to \eqref{may15-1}, where $x_0 =\eta$ is now {\it conformal}
time~\cite{vrscalinv}. The Lagrangian for $\Theta$ coincides with the
second term in \eqref{may17-2}, while the perturbations $\delta \rho$ are
again governed by Eq.~\eqref{may18-21}. The properly normalized solutions
for the field $\theta$ and perturbations $\delta \rho$ coincide with
\eqref{may18-1} and \eqref{apr26-3}, respectively, with  $h$ being now the
quartic self-coupling  and $x_0 =\eta$. So, the dynamics of
perturbations in the conformal rolling scenario is identical to that in
the Galilean Genesis.

\subsection{General argument}
\label{sec:general}

Let us see that the form of Eq.~\eqref{may18-21} that governs the
 perturbations $\delta \rho$ at the linearized level is completely determined
 by
conformal invariance. Namely, let us consider any weakly coupled theory,
in which the classical field equation for $\rho$ is second order in
derivatives, the action for $\rho$ is local, invariant under space-time
translations and spatial rotations and invariant under scaling
\be
\rho (x) \to \lambda \rho(\lambda x) \nonumber
\ee
and inversion
\be
\rho (x^\mu) \to \frac{1}{x^2} \ \rho \l \frac{x^\mu}{x^2} \r \;.
\nonumber
\ee
One designs a theory in such a way that it admits the runaway solution
$\rho_c = - x_0^{-1}$  (one can always set the overall constant
in $\rho_c$ equal to 1 by field redefinition).
This solution is invariant
under both scaling and inversion. So, the quadratic action
for perturbations $\delta \rho$ about this solution
must also be invariant. Let us write for
the quadratic action
\be
S^{(2)} = h^{-2} \int~d^4x~\delta \rho \ {\cal L} \ \delta \rho \; ,
\label{may24-2}
\ee
where ${\cal L}$ is second order differential operator, and $h$ is some
constant whose choice is specified below. Since the background depends
only on time, the operator ${\cal L}$ does not contain spatial coordinates
explicitly, but may contain time. Also, it is invariant under spatial
rotations.

Invariance under scaling implies that modulo overall constant, the part of
${\cal L}$ that involves derivatives is
\be
{\cal L} \supset - \d_0^2 + v_s^2 \Delta \; .
\ee
where $v_s$ is independent of time. We can choose the constant $h$ in
\eqref{may24-2} in such a way that the term with two time derivatives
enters with the coefficient $-1$. Scale invariance alone is
insufficient\footnote{A straightforward way to see this is to consider a
modification of the galileon theory in which the second and third terms in
\eqref{may24-3} enter with unrelated coefficients.} for obtaining $v_s=1$.
However, the requirement of invariance under inversion uniquely specifies
$v_s=1$. The term without derivatives is almost uniquely determined from
the requirement of invariance under scaling and inversion,
\be
{\cal L} \supset \frac{c}{x_0^2}
\ee
where $c$ is yet undetermined constant. To complete the argument we
notice that the invariance of the original action under time translations
implies that $\delta \rho = \dot{\rho}_c = x_0^{-2}$ must be a solution to
equation ${\cal L} \delta \rho = 0$. This gives $c=6$. Thus, the whole quadratic
action for perturbations is uniquely determined by conformal invariance,
modulo an overall constant factor, and the linearized equation for $\delta
\rho$ has one and the same form in the whole class of
conformally-invariant models with the Lagrangians of the general
form \eqref{may17-2}. To the leading {\it non-linear} order, the properties
of the $\theta$-perturbations are identical in these models, as they are
governed by the interaction Hamiltonian \eqref{Eq/Pg3/3:in-in}.

\section{Trispectrum}
\label{sec:main}

\subsection{Generalities}

In models with the Lagrangians of the form \eqref{may17-2},
the bispectrum of the field
$\theta$ vanishes. To calculate the trispectrum we make use of the in-in
formalism, cf. Ref.~\cite{Maldacena:2002vr}. We use the shorthand notation
\[
\theta^4_{\bf x} \equiv \theta (x) \theta (y) \theta (z) \theta (w) \; ,
\]
with understanding that we are interested in the formal limit
\be
x_0 = y_0 =z_0 =w_0 \to 0 \; .
\label{may18-51}
\ee
Then the four-point function reads
\[
\langle \theta^4_{\bf x}
 \rangle =\left\langle \left[\overline{T}\exp\left(i\int
\limits_{-\infty }^{0}dx_{0}H_{I} \right)\right] \theta_{(I)}^4
\left[T\exp\left(-i\int \limits_{-\infty }^{0}dx_{0}H_{I}
\right)\right]\right\rangle
\]
where subscript $I$ refers to interaction picture and the interaction
Hamiltonian density is given by \eqref{Eq/Pg3/3:in-in}. Of course, we are
going to calculate the connected part of the four-point function.

To this end we need the two-point functions of the linear fields
 $\theta_{(I)}$ and $\delta \rho_{(I)}$,
\[
\langle \theta_{(I)} (x) \theta_{(I)} (y) \rangle \equiv D(x,y) \; ,
\;\;\;\;\;\; \langle \delta \rho_{(I)} (x) \delta \rho_{(I)} (y) \rangle
\equiv \frac{h^2}{2} D_{\rho }(x,y)\;.
\]
By making use of \eqref{may18-1} and \eqref{apr26-3} we find
\begin{eqnarray}
D(x,y)&=&\frac{1}{32\pi ^{2}}\int ~d^{3}k~ (x_0 y_0)^{3/2}
~H_{3/2}^{(1)}(-kx_{0})H_{3/2}^{(2)}(-ky_{0})
\mathrm{e}^{i\mathbf{k}(\mathbf{x}-\mathbf{y})}
\label{Eqn/Pg4/1:in-in}\\
D_{\rho }(x,y)&=&\frac{1}{32\pi ^{2}}\int ~d^{3}k~ (x_0 y_0)^{1/2}~
H_{5/2}^{(1)}(-kx_{0})H_{5/2}^{(2)}(-ky_{0})
\mathrm{e}^{i\mathbf{k}(\mathbf{x}-\mathbf{y})}\;.
\label{Eqn/Pg4/2:in-in}
\end{eqnarray}
Both of these pairing functions satisfy
\[
D_{(\rho )}(x,y)=D_{(\rho )}^{*}(y,x)\;.
\]
We also need the (anti)$T$-product of the field $\delta \rho $. We
write
\begin{eqnarray}
\overline{T}_{xy}&+&T_{xy}\equiv \frac{2}{h^2}\left[ \langle
\overline{T}(\delta \rho_{(I)}(x) \delta \rho_{(I)}(y)) \rangle + \langle
{T}(\delta \rho_{(I)}(x) \delta \rho_{(I)}(y)) \rangle \right]
=\nonumber\\
&=&\Theta (x_{0}-y_{0})D_{\rho }(y,x)+ \Theta (y_{0}-x_{0})D_{\rho
}(x,y)
+ \Theta (x_{0}-y_{0})D_{\rho }(x,y)+ \Theta (y_{0}-x_{0})D_{\rho }(y,x)
\nonumber\\
&=&D_{\rho }(x,y)+D_{\rho }(y,x)=2 \mathrm{Re}[D_{\rho }(x,y)]
\label{Eq/Pg4/3:in-in}
\end{eqnarray}
and
\begin{equation}
\overline{T}_{xy}-T_{xy}=-\varepsilon (x_{0}-y_{0})[D_{\rho }(x,y)-D_{\rho
}(y,x)]=-2i\varepsilon (x_{0}-y_{0})\mathrm{Im}[D_{\rho }(x,y)] \; .
\label{Eq/Pg4/4:in-in}
\end{equation}
The  connected four-point function is a sum of the three terms,
one of which has the following form,
\begin{eqnarray}
&& \left\langle \left( i  \int \limits_{}^{}d^{4}x'\mathcal{
H}_{I}(x')\right) \theta_{(I)} ^{4} \left(-i \int
\limits_{}^{}d^{4}x''\mathcal{ H}_{I}(x'')\right)\right\rangle \nonumber\\
&&=
2h^{2}\int \limits_{}^{}\frac{d^{4}x'd^{4}x''}{x^{\prime}_0 x_0^{\prime
\prime}}\left[\rule{0pt}{6mm}\partial _{\mu }'D(x',x)\partial _{\mu
}'D(x',y) \partial _{\nu }''D(z,x'')\partial _{\nu }''D(w,x'')D_{\rho
}(x',x'') +\right.
\nonumber\\
&&+
\left.\left(
\begin{array}{c}
x \leftrightarrow z\\
y \leftrightarrow w\\
\end{array}
\right) \right]
+ \{ y \leftrightarrow z\}+ \{ y \leftrightarrow w\}
\label{Eq/Pg5/1:in-in}
\end{eqnarray}
and the two others are
\begin{eqnarray}
&&\left \langle \left( \frac{i^{2}}{2} \int \limits_{}^{}d^{4}x'd^{4}x''
\overline{T}\mathcal{ H}_{I}(x')\mathcal{ H}_{I}(x'')\right) \theta_{(I)}
^{4}\right\rangle+ \left \langle \theta_{(I)} ^{4}\left(
\frac{(-i)^{2}}{2}  \int \limits_{}^{}d^{4}x'd^{4}x'' T\mathcal{
H}_{I}(x')\mathcal{ H}_{I}(x'')\right)\right \rangle \nonumber\\
&&=
-2h^{2}\int \limits_{}^{}\frac{d^{4}x'd^{4}x''}{x^{\prime}_0 x_0^{\prime
\prime}}\left[\partial _{\mu }'D(x',x)\partial _{\mu }'D(x',y) \partial
_{\nu }''D(x'',z)\partial _{\nu }''D(x'',w)\overline{T}_{x'x''} +
\phantom{\frac{\phantom{1}}{\phantom{1}}} \right.
\nonumber\\
&&+
\left. \phantom{\frac{\phantom{1}}{\phantom{1}}} \partial _{\mu
}'D(x,x')\partial _{\mu }'D(y,x') \partial _{\nu }''D(z,x'')\partial _{\nu
}''D(w,x'')T_{x'x''} \right]\rule{0pt}{6mm}
+ \{ y \leftrightarrow z\}+ \{ y \leftrightarrow w\}\;.
\label{Eq/Pg5/125:in-in}
\end{eqnarray}
In the 3-dimensional momentum representation, these expressions reduce to
the integrals over $dx_0^\prime$ and $dx_0^{\prime \prime}$. Our
definition of the $n$-point function in the momentum representation is
\[
\langle \prod \limits_{i=1}^{n}\theta _{\mathbf{k}_{i}}\rangle =(2\pi
)^{3}\delta \left(\sum
\limits_{i=1}^{n}\mathbf{k}_{i}\right)G_{n}(\mathbf{k}_{1},\ldots,\mathbf{k}_{n})
=\int \limits_{}^{}\prod
\limits_{i=1}^{n}\left\{d^{3}x_{i}\mathrm{e}^{-i\mathbf{k}_{i}\mathbf{x}_{i}}
\right\}G_{n}(\mathbf{x}_{1},\ldots,\mathbf{x}_{n}),
\]
which corresponds to
\[
\theta (\mathbf{k})=\int \limits_{}^{}d^{3}x
\mathrm{e}^{-i\mathbf{kx}}\theta (x) \; .
\]
With our convention, the power spectrum is related to the two-point
function by
\[
\langle \theta (\mathbf{k})\theta (\mathbf{k}^\prime)\rangle =(2\pi
)^{5}\delta (\mathbf{k}+\mathbf{k}^\prime)\frac{1}{2k^{3}}\mathcal{
P}_{\theta } (k)
\]
and at the linearized level $\mathcal{ P}_{\theta }$ is given by
\eqref{may18-50}.

\subsection{Leading singularity}

Let us calculate the contribution to the trispectrum due to the term
explicitly written in square brackets in \eqref{Eq/Pg5/1:in-in}. In the
limit \eqref{may18-51} it involves the following combination
\begin{eqnarray}
&&  \partial _{\mu }'D(x',\mathbf{x})\partial _{\mu
}'D(x',\mathbf{y})=-\frac{1}{(32\pi ^{2})^{2}}\frac{2}{\pi }\int
\limits_{}^{}\frac{d^{3}k_{1}d^{3}k_{2}}{k_{1}^{3/2}k_{2}^{3/2}}
\mathrm{e}^{i\mathbf{k_{1}(x'-x)}+i\mathbf{k_{2}(x'-y)}} \nonumber\\
&& \times \left[\frac{\partial }{\partial \xi }\left(\xi
^{3/2}H_{3/2}^{(1)}(k_{1}\xi )\right)\frac{\partial }{\partial \xi
}\left(\xi ^{3/2}H_{3/2}^{(1)}(k_{2}\xi
)\right)+\mathbf{k}_{1}\mathbf{k}_{2}\xi ^{3}H_{3/2}^{(1)}(k_{1}\xi
)H_{3/2}^{(1)}(k_{2}\xi ) \right]
\nonumber \\
&& = -\frac{1}{(32\pi ^{2})^{2}}\frac{2}{\pi^2 }\int
\limits_{}^{}\frac{d^{3}k_{1}d^{3}k_{2}}{k_{1}^{3}k_{2}^{3}}
\mathrm{e}^{i\mathbf{k_{1}(x'-x)}+i\mathbf{k_{2}(x'-y)}}
\left[\tilde{R}(k_{1},k_{2},k_{12},\xi ) + i
\tilde{Q}(k_{1},k_{2},k_{12},\xi ) \right] \; ,
\end{eqnarray}
where
\[
\xi = - x_0^\prime \; ,
\]
\begin{eqnarray}
\tilde{R}(k_{1},k_{2},k_{12},\xi )&=& \cos(\xi
(k_{1}+k_{2}))\cdot\left[\xi ^{2}k_{1}k_{2}(k_{12}^{2}-(k_{1}+k_{2})^{2})
+k_{1}^{2}+k_{2}^{2}-k_{12}^{2}\right] \nonumber\\
&+&\sin(\xi (k_{1}+k_{2}))\cdot\xi
(k_{1}+k_{2})(k_{1}^{2}+k_{2}^{2}-k_{12}^{2})
\label{Eq/Pg8/1:in-in}
\end{eqnarray}
and
\begin{eqnarray}
\tilde{Q}(k_{1},k_{2},k_{12},\xi )&=& \sin(\xi
(k_{1}+k_{2}))\cdot\left[\xi ^{2}k_{1}k_{2}(k_{12}^{2}-(k_{1}+k_{2})^{2})
+k_{1}^{2}+k_{2}^{2}-k_{12}^{2}\right]\nonumber\\
&-&\cos(\xi (k_{1}+k_{2}))\cdot\xi
(k_{1}+k_{2})(k_{1}^{2}+k_{2}^{2}-k_{12}^{2})\; .
\label{Eqn/Pg8/1:in-in}
\end{eqnarray}
with
$\mathbf{k_{12}=k}_{1}+\mathbf{k}_{2}$.
So, we obtain for the contribution under study (in a certain sense this is
the leading contribution, hence the notation)
\begin{eqnarray}
&&G
_{4(l.c.)}(\mathbf{k}_{1},\mathbf{k}_{2},\mathbf{k}_{3},\mathbf{k}_{4})=
\frac{ \pi h^{2} }{32}\frac{1}{k_{1}^{3}k_{2}^{3}k_3^{3}k_4^{3}}
\nonumber \\
&&\phantom{\times }\times \!\!\int \limits_{0}^{\infty }\!\!\frac{d\xi
d\chi}{\sqrt{\xi \chi}} \tilde{Q}(k_{1},k_{2},k_{12},\xi
)\tilde{Q}(k_3,k_4,k_{12},\chi )\left[Y_{5/2}(k_{12}\xi
)Y_{5/2}(k_{12}\chi )+J_{5/2}(k_{12}\xi )J_{5/2}(k_{12}\chi )\right] \; ,
\nonumber
\end{eqnarray}
where
\[
\chi=-x^{\prime \prime}_0 \;
\]
and we made use of the fact that $\mathbf{k}_{3}+\mathbf{k}_{4} = -
\mathbf{k}_{12}$. To regularize integrals at infinity we insert
$\exp(-\varepsilon \xi ) \exp(-\varepsilon \chi )$ into the integrand and
take the limit $\varepsilon \to +0$ in the end of the calculation. This is
equivalent to the replacement $x_{0}\to x_{0}(1-i\varepsilon) $ which is
the standard prescription for calculating vacuum expectation values in the
interaction picture (see, e.g., Ref.~\cite{Maldacena:2002vr}). At $\xi \to
0$ all integrals are converging. We observe that the integrals over $\xi$
and $\chi$ factor out and obtain
\be
G_{4(l.c.)}(\mathbf{k}_{1},\mathbf{k}_{2},\mathbf{k}_{3},\mathbf{k}_{4})=
\frac{ \pi h^{2} }{32}\frac{1}{k_{1}^{3}k_{2}^{3}k_3^{3}k_4^{3}} \left[
\mathcal{ Y}(k_{1},k_{2},k_{12}) \mathcal{ Y}(k_3,k_4,k_{12}) + \mathcal{
J}(k_{1},k_{2},k_{12}) \mathcal{ J}(k_3,k_4,k_{12}) \right] \; ,
\label{may26-10}
\ee
where
\begin{eqnarray}
\mathcal{ Y}(k_{1},k_{2},k_{12})&\equiv&\int \limits_{0}^{\infty
}\frac{d\xi}{\sqrt{\xi}} \tilde{Q}(k_{1},k_{2},k_{12},\xi
)Y_{5/2}(k_{12}\xi )\nonumber\\
&=&-\frac{1}{2}\sqrt{\frac{\pi
}{2}}\frac{1}{k_{12}^{5/2}}\left[3(k_{1}^{2}-k_{2}^{2})^{2} -
2(k_{1}^{2}+k_{2}^{2})k_{12}^{2}-k_{12}^{4}\right]
\label{Eqn/Pg9/2:in-in}
\end{eqnarray}
and
\begin{eqnarray}
\mathcal{ J}(k_{1},k_{2},k_{12})&\equiv&\int \limits_{0}^{\infty
}\frac{d\xi}{\sqrt{\xi}} \tilde{Q}(k_{1},k_{2},k_{12},\xi
)J_{5/2}(k_{12}\xi )\nonumber\\
&=&\frac{1}{\sqrt{2\pi }k_{12}^{5/2}}
\left[\rule{0pt}{5mm}k_{12}(k_{12}^{2}-3(k_{1}-k_{2})^{2})(k_{1}+k_{2})
\phantom{\frac{1}{1}}\right.\nonumber\\
&+&\left.\left(3(k_{1}^{2}-k_{2}^{2})^{2} -
2(k_{1}^{2}+k_{2}^{2})k_{12}^{2}-k_{12}^{4}\right)
\mathrm{arctanh}\frac{k_{12}}{k_{1}+k_{2}} \right]\;.
\label{Eqn/Pg9/3:in-in}
\end{eqnarray}
The expression \eqref{Eqn/Pg9/2:in-in} is singular in the folded limit
$k_{12} \to 0$, while $\mathcal{ J}(k_{1},k_{2},k_{12})$ vanishes in this
limit. The crossing terms in \eqref{Eq/Pg5/1:in-in}, as well as the terms
\eqref{Eq/Pg5/125:in-in} are regular as  $k_{12} \to 0$ (the latter
property is established by direct calculation in  Appendix). It is
convenient to define the singular part of $\mathcal{ Y}$ in the following
way,
\be
\mathcal{ Y}_{s}(k_{1},k_{2},k_{12})=-\frac{1}{2}\sqrt{\frac{\pi
}{2}}\frac{1}{k_{12}^{5/2}}\left[3(k_{1}^{2}-k_{2}^{2})^{2} -
2(k_{1}^{2}+k_{2}^{2})k_{12}^{2}+k_{12}^{4}\right] \; , \nonumber
\ee
so that
\[
\mathcal{ Y} = \mathcal{ Y}_s + \mathcal{ Y}_r \; , \;\;\;\;\;
 \mathcal{ Y}_r =
\sqrt{\frac{\pi }{2}}k_{12}^{3/2} \; .
\]
 Then the singular part of the four-point function is
\be
G_{4(s)}(\mathbf{k}_{1},\mathbf{k}_{2},\mathbf{k}_{3},\mathbf{k}_{4})=
\frac{ \pi h^{2} }{32}\frac{1}{k_{1}^{3}k_{2}^{3}k_3^{3}k_4^{3}} \mathcal{
Y}_s(k_{1},k_{2},k_{12}) \mathcal{ Y}_s(k_3,k_4,k_{12}) ~~~ +~
2~~\mbox{permutations}
\label{Eq/Pg12/2:in-in}
\ee
In the limit $k_{12} \to 0$ we recover the result \eqref{may21-1},
which can be written in a symmetric form,
\be
G_{4(s)}(\mathbf{k}_{1},\mathbf{k}_{2},\mathbf{k}_{3},\mathbf{k}_{4})
\vert_{k_{12} \to 0}
= \frac{\pi^2 h^{2}}{16} \frac{1}{k_{12}Q^{4}P^{4}} \left[1-3\left(
\frac{\mathbf{k_{12} Q}}{k_{12} Q} \right)^{2} \right]
\left[1-3\left(\frac{\mathbf{k_{12} P}}{k_{12} P} \right)^{2} \right] \; ,
\nonumber
\ee
where
\[
\mathbf{Q} = \frac{\mathbf{k}_1 - \mathbf{k_2}}{2} \; , \;\;\;\;\;\;\;
\mathbf{P} = \frac{\mathbf{k}_3 - \mathbf{k_4}}{2} \; .
\]
Note that the terms of order  $1/k_{12}^5$ and $1/k_{12}^3$, which one
 could off hand expect from (\ref{Eqn/Pg9/2:in-in}), cancel out. Thus, the
field $\theta$ has rather mild infrared behavior, even though it interacts
with infrared-enhanced modes of $\delta \rho$ (the power spectrum of
$\delta \rho$ is red, see \eqref{may18-20}). The reason for this property
is discussed in Ref.~\cite{Libanov:2010nk}.

\subsection{Shapes}
\label{sec:shapes}
The calculation of the contribution
\eqref{Eq/Pg5/125:in-in} is not so straightforward. We perform this
calculation in  Appendix, where we also present the complete result for the
trispectrum. The definition of the trispectrum $\mathcal{T}$ in our
case is
\[
G_{4}(\mathbf{k}_{1},\mathbf{k}_{2},\mathbf{k}_{3},\mathbf{k}_{4}) =
\frac{h^{2}}{\prod \limits_{i=1}^{4}k_{i}^{3}}\mathcal{
T}(k_{1},k_{2},k_{3},k_{4},k_{12},k_{14})\;,
\]
where $\mathbf{k}_{14} = \mathbf{k}_{1} + \mathbf{k}_{4}$.
The combinations $k_{13}  =|\mathbf{k}_{1} + \mathbf{k}_{3}|$ and
$k_{24} =| \mathbf{k}_{2} + \mathbf{k}_{4}|$ are not independent,
since
\[
k_{13}= k_{24} =
\sqrt{k_{1}^{2}+k_{2}^{2}+k_{3}^{2}+ k_{4}^{2}-k_{12}^{2}-k_{14}^{2}} \;.
\]
It is
convenient for the purpose of illustration to decompose the trispectrum
into the part  $\mathcal{T}_s$, which is singular as either $k_{12} \to 0$
or  $k_{13} \to 0$ or  $k_{14} \to 0$, and the regular part $\mathcal{T}_r$,
\[
\mathcal{ T}=\mathcal{ T}_{s}+\mathcal{ T}_{r} \; .
\]
The singular part is obtained\footnote{The
product $\mathcal{ Y}_{s}\mathcal{ Y}_{r}$ is regular in the limit $k_{12}
\to 0$.} from \eqref{Eq/Pg12/2:in-in},
\be
\mathcal{T}_s = \frac{\pi}{32} \left[ \mathcal{
Y}_s(k_{1},k_{2},k_{12}) \mathcal{ Y}_s(k_3,k_4,k_{12}) + (\mathbf{k}_2
\leftrightarrow \mathbf{k}_3) + (\mathbf{k}_2 \leftrightarrow
\mathbf{k}_4) \right] \; ,
\label{may19-1}
 \ee
 while the rest can be read off from \eqref{Eq/Pg15/2:in-in} and
 \eqref{Eq/Pg16/1:in-in}. We emphasize that there are no  other
singularities in $\mathcal{ T}$: one can check that the logarithmic
divergences appearing in intermediate formulas cancel out when one takes
 into account all contributions.

Following Ref.~\cite{Chen:2009bc},  we are going to
compare our trispectrum to the trispectra of the local forms. The latter
are obtained from the Ansatz in real
space~\cite{Okamoto:2002ik,Kogo:2006kh}
\be
\zeta (\mathbf{x})=\zeta _{g}+\frac{3}{5}f_{NL}(\zeta _{g}^{2}-\langle
\zeta _{g}^{2}\rangle )+\frac{9}{25}g_{NL}(\zeta _{g}^{3}-3\langle \zeta
_{g}^{2}\rangle \zeta _{g}) \; ,
\label{may24-1}
\ee
where $\zeta_g$ is the Gaussian field and $f_{NL}$ and
$g_{NL}$ are constants. One has~\cite{Chen:2009bc}
\[
G_{4,
\mathrm{loc}}(\mathbf{k}_{1},\mathbf{k}_{2},\mathbf{k}_{3},\mathbf{k}_{4})
= \frac{\mbox{const}}{\prod \limits_{i=1}^{4}k_{i}^{3}} \l f_{NL}^{2}\mathcal{
T}_{\mathrm{loc1}}+ g_{NL}\mathcal{ T}_{\mathrm{loc2}} \r \;,
\]
where the two local shapes are
\begin{eqnarray}
\mathcal{ T}_{\mathrm{loc1}}&=&\frac{9}{50}\left(
\frac{k_{1}^{3}k_{3}^{3}+k_{1}^{3}k_{4}^{3}+
k_{2}^{3}k_{3}^{3}+k_{2}^{3}k_{4}^{3}}{k_{12}^{3}}
+\{\mathbf{k}_{2}\leftrightarrow \mathbf{k}_{3}\}
+\{\mathbf{k}_{2}\leftrightarrow \mathbf{k}_{4}\} \right) \;,
\label{Eqn/Pg18/1A:in-in}\\
\mathcal{ T}_{\mathrm{loc2}}&=&\frac{27}{100}\sum
\limits_{i=1}^{4}k_{i}^{3}.
\label{Eqn/Pg18/1:in-in}
\end{eqnarray}
Let us quantify the strength of the non-Gaussianity. The standard
estimator $t_{NL}$ is related to the four-point function in the regular
tetrahedron limit, $k_{i}=k_{12}=k_{14}\equiv k$,
\be
\langle \zeta_{\mathbf{k}}^{4}\rangle_{k_{i}=k_{12}=k_{14}\equiv k}
=(2\pi )^{9}\mathcal{ P}_{\zeta }^{3}\delta \left(\sum
\limits_{i=1}^{4}\mathbf{k}_{i}   \right)\frac{1}{k^{9}}t_{NL} \; .
\label{aug31-1}
\ee
Note that the standard definition involves $\mathcal{ P}_{\zeta }^{3}$
in the right hand side. This is appropriate for the local Ansatz
\eqref{may24-1}, and the sizes in the local models are~\cite{Chen:2009bc}
\[
t_{NL}^{\mathrm{loc1}}=2.16f_{NL}^{2},\ \ \
t_{NL}^{\mathrm{loc2}}=1.08g_{NL}.
\]
On the other hand, the quantity we can directly calculate in our model
is
\[
\langle \theta_{\mathbf{k}}^{4}\rangle_{k_{i}=k_{12}=k_{14}\equiv k}
=(2\pi )^{9}\mathcal{ P}_{\theta }^{3}\delta \left(\sum
\limits_{i=1}^{4}\mathbf{k}_{i}   \right)\frac{1}{k^{9}}t^{(\theta)}_{NL}
 \; ,
\]
where $\mathcal{ P}_{\theta } = 1/(4\pi^2)$, see \eqref{may18-50}, and
the size of the non-Gaussianity in $\theta$ in our model is obtained from
\eqref{Eq/Pg15/2:in-in} and \eqref{Eq/Pg16/1:in-in},
\be
t^{(\theta)}_{NL}=2.87  h^{2} \; .
\label{aug31-2}
\ee
The adiabatic perturbation $\zeta$ is proportional to $\theta$,
namely,
\be
\zeta = r \frac{\theta}{\bar{\Theta}} \; ,
\label{sep1-1}
\ee
where $\bar{\Theta}$ is the homogeneous background value of the
scalar field, and $r\lesssim 1$ is a dilution factor, which is
 independent of $k$ for
both curvaton and modulated decay mechanism of conversion of the
$\theta$-perturbations into adiabatic ones.
Therefore, in our model the size of the non-Gaussianity of the adiabatic
perturbations is
\[
 t_{NL} = \frac{{\cal P}_\theta}{{\cal P}_\zeta} \cdot t_{NL}^{(\theta)}
 = 2.87 \frac{h^2}{4\pi^2 {\cal P}_\zeta} \; .
 \]
As discussed in Section~\ref{sec:Intro}, there are no model-independent
constraints on $\bar{\Theta}$ and $h$, so the intrinsic non-Gaussianity
we study in this paper may well dominate over conversion-related
one and be detectable.

Let us point out, however, that in the conformal rolling scenario
of Ref.~\cite{vrscalinv}, the adiabatic power spectrum is
itself proportional to $h^2$. The reason is that $\Theta$ is the
phase field. With our normalization, its background value is
bounded from above, see Eq.~(\ref{sep1-2}),
\[
|\bar{\Theta}| \leq \frac{\sqrt{2} \pi}{h} \; ,
\]
and without fine tuning $|\bar{\Theta}| \sim \pi/h$. Equation
(\ref{sep1-1}) then gives
\[
{\cal P}_\zeta \geq \frac{r^2 h^2}{8 \pi^4} \; ,
\]
so that
\[
t_{NL} \leq 2.87 \frac{2\pi^{2}}{r^2} \; .
\]
Therefore, the intrinsic non-Gaussianity can be sizeable
only for small dilution factor $r$. If the $\theta$-perturbations
are converted into adiabatic ones by the curvaton mechanism,
the conversion-related non-Gaussianity is large at small $r$,
$f_{NL} \sim r^{-1}$~\cite{Lyth:2002my}, so its contribution to the
trispectrum is very roughly of the same order as that due to
the intrinsic non-Gaussianity. The analysis of the detectability
of the intrinsic non-Gaussianity in this case deserves further study.
The same remark applies to the bulk of the modulated decay models,
where $f_{NL} \sim r^{-1}$ as well~\cite{Zaldarriaga:2003my}. There is an
exception, however~\cite{Ichikawa:2008ne}: if the width of decaying particles
$\Gamma (\theta)$ is linear in $\theta$ or has the form
$\Gamma (\theta) = (\gamma_0 + \gamma_1 \theta)^2$ (which is more
plausible from particle physics prospective), then $f_{NL}$,
$g_{NL}$ are roughly of order 1 even for small
$r \sim \gamma_1/\gamma_0$, and hence large $t_{NL}$.
We conclude that in the conformal rolling scenario,
the domination and detectability of the intrinsic non-Gaussianity
is possible, but not at all generic.

Let us now turn to the shapes.
To compare them, we set in what follows
\[
\frac{h^2}{4\pi^2 {\cal P}_\zeta}
=f_{NL}=g_{NL} =1 \; ;
\]
then
the sizes of the trispectra in all models are similar.

\begin{figure}
\includegraphics[width=\textwidth]{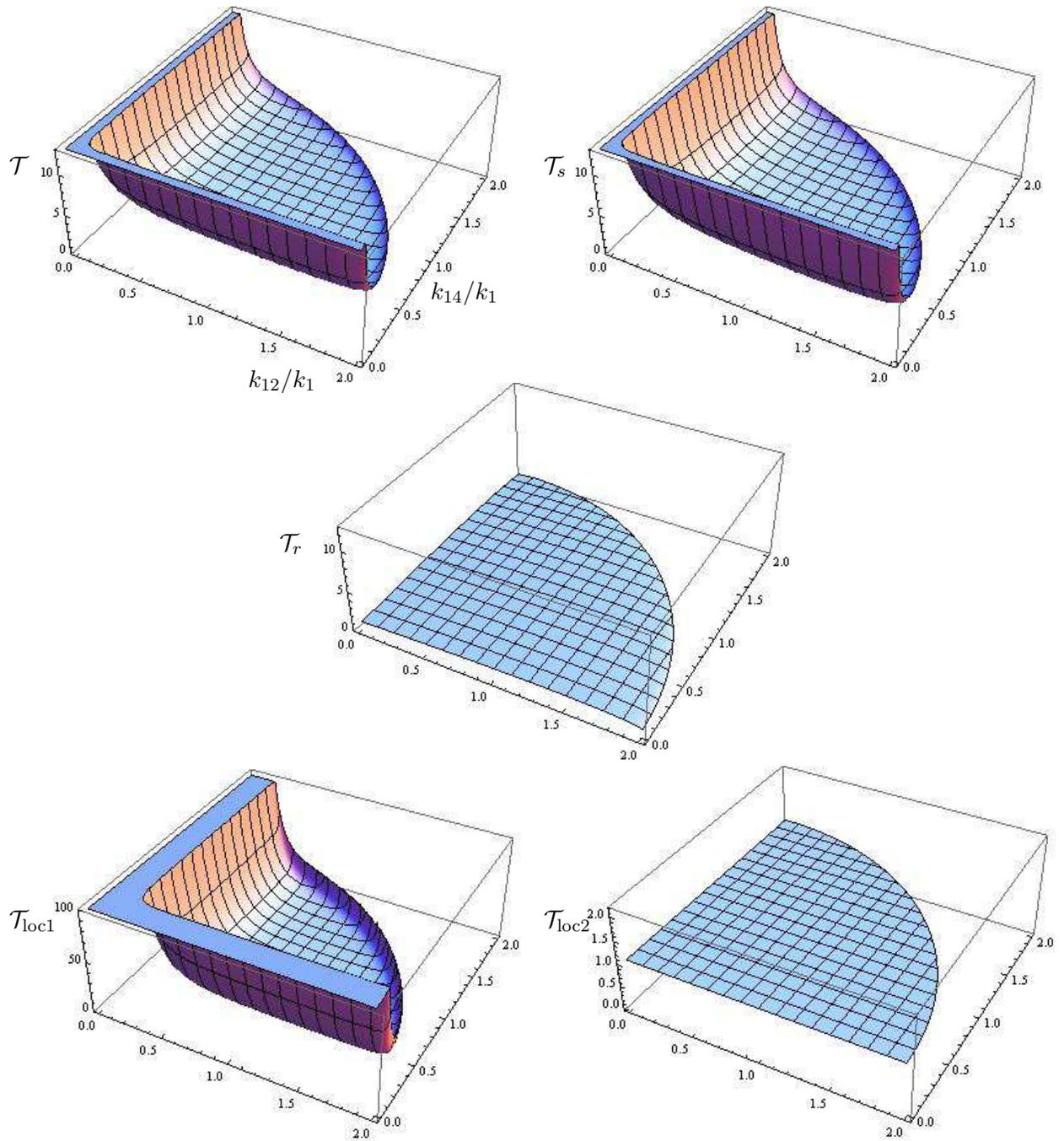}
\caption{
\label{F:eq}
Complete trispectrum $\mathcal{T}$ (upper left panel), its
singular part $\mathcal{T}_s$ (upper right panel), its regular part
$\mathcal{T}_r$ (middle panel), trispectrum of local form $\mathcal{
T}_{\mathrm{loc1}}$ (lower left panel) and trispectrum of another local
form  $\mathcal{ T}_{\mathrm{loc2}}$ (lower right panel) in
\textit{equilateral limit.} See Section~\ref{sec:shapes} for definitions.
Note that vertical scales in the lower panels are different, and neither
coincides with the vertical scales in the three upper panels.}
\end{figure}

\begin{figure}
\includegraphics[width=\textwidth]{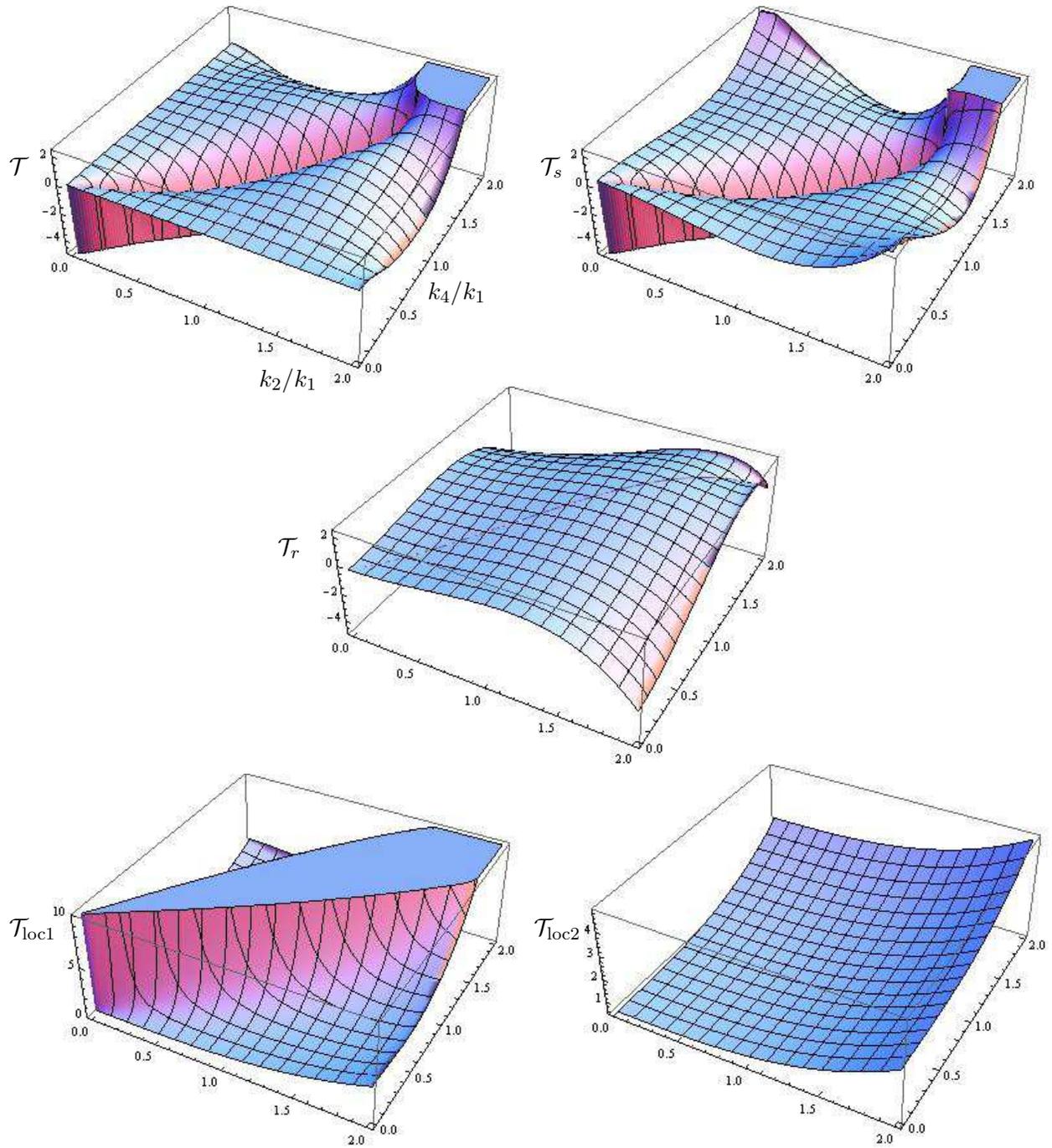}
\caption{
\label{F:SPL}
Same as in Fig.~\ref{F:eq}, but in \textit{specialized
planar limit.} }
\end{figure}

\begin{figure}
\includegraphics[width=\textwidth]{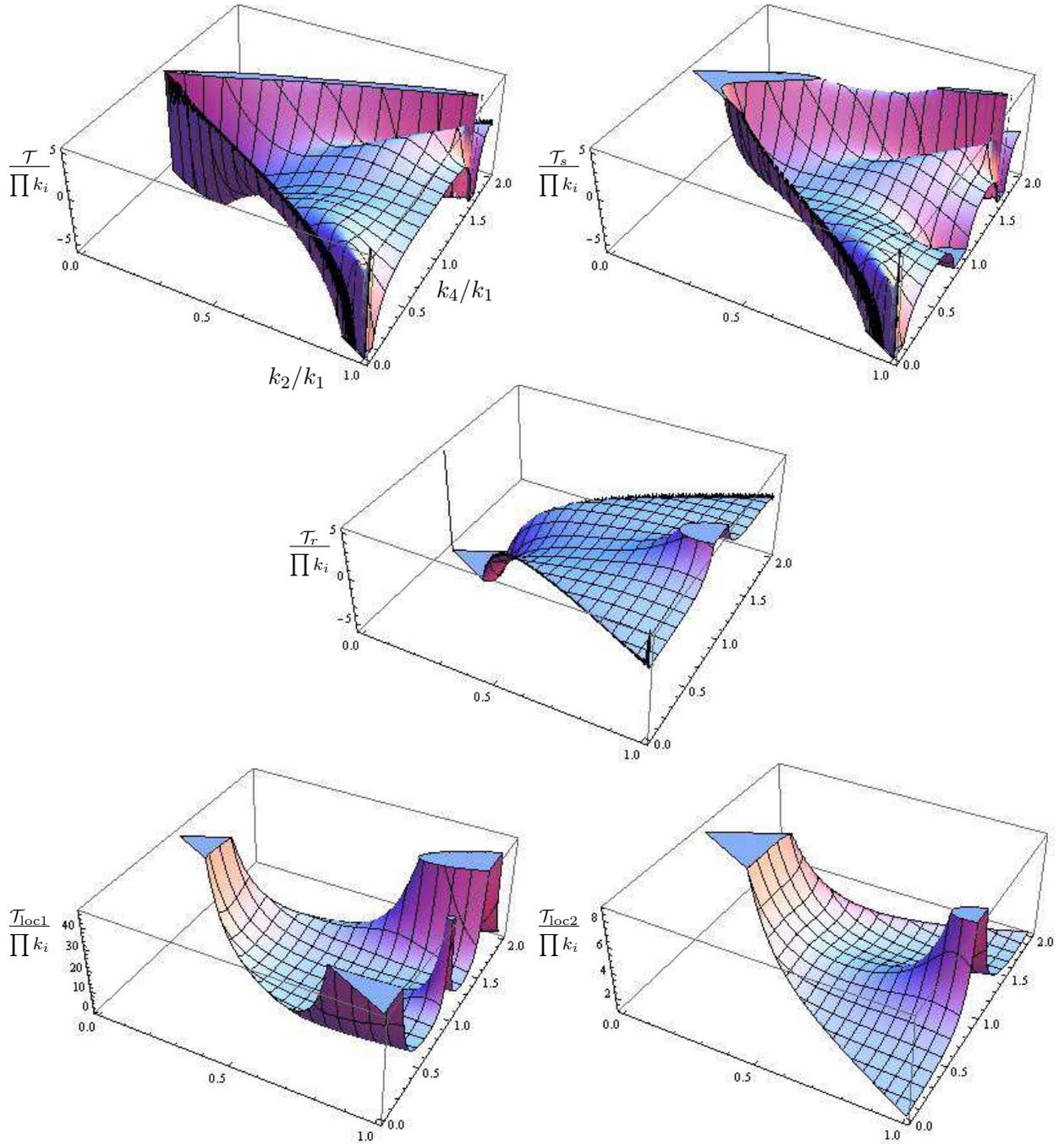}
\caption{
\label{F:NDSL}
\textit{Near the double-squeezed limit.} Shown are the
combinations $\mathcal{ T}/(\prod \limits_{i=1}^{4}k_{i})$ for the same
trispectra as in Fig.~\ref{F:eq}.
}
\end{figure}

\begin{figure}
\includegraphics[width=\textwidth]{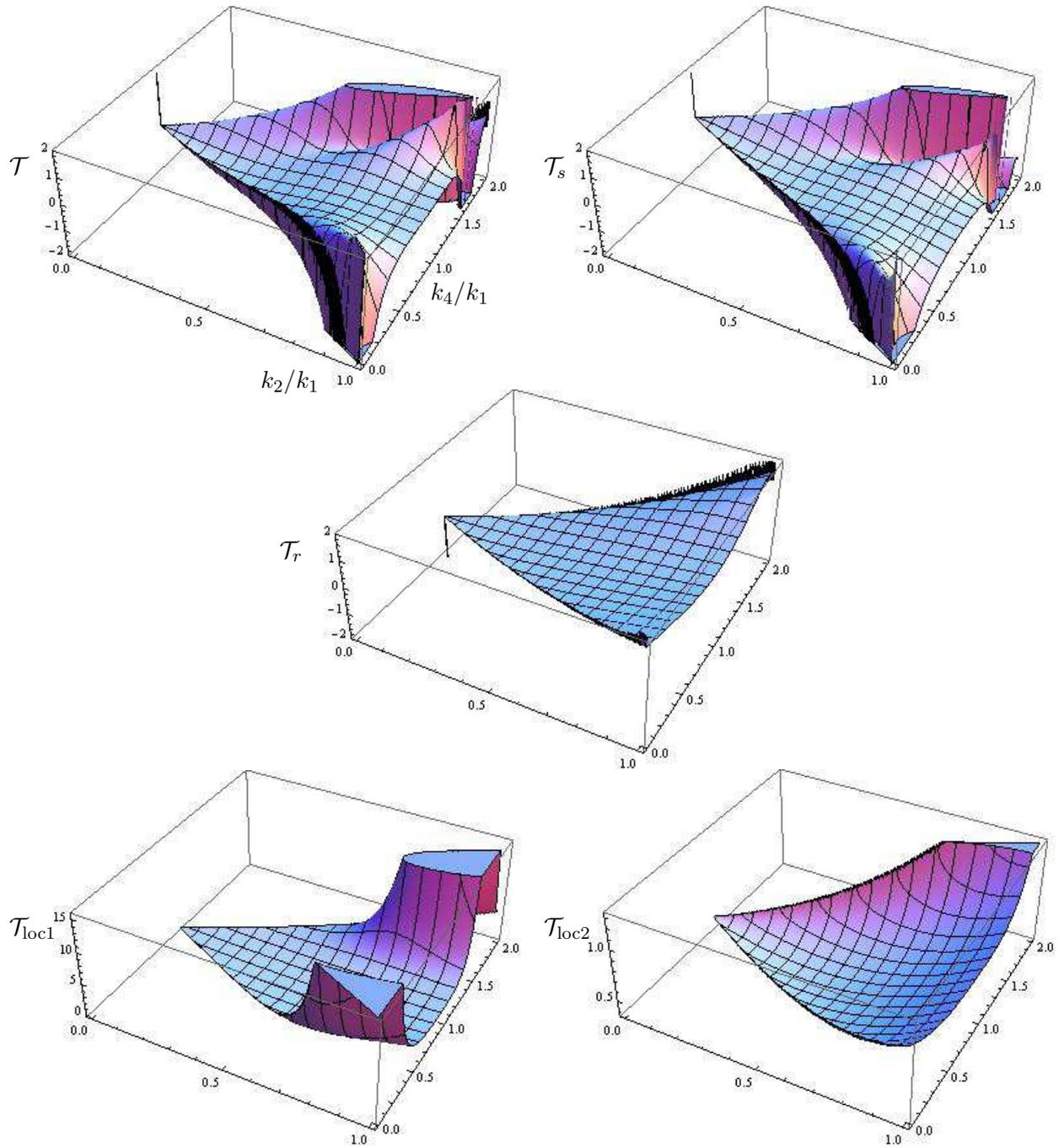}
\caption{
\label{F:NDSL-2}
\textit{Near the double-squeezed limit.} Same as in
Fig.~\ref{F:NDSL}, but for the trispectra $\mathcal{ T}$ themselves. }
\end{figure}

Let us consider various limits of the shape function $\mathcal{T}$. We use
the nomenclature of Refs.~\cite{Chen:2009bc,Bartolo:2010di}. The first
three panels in Figs.~\ref{F:eq}, \ref{F:SPL}, \ref{F:NDSL} and
\ref{F:NDSL-2} show the complete trispectrum $\mathcal{T}$, the
contribution of the singular part $\mathcal{T}_s$ and the regular part
$\mathcal{T}_r$ in our model, respectively. The fourth and fifth panels
show the two local trispectra $\mathcal{ T}_{\mathrm{loc1}}$ and
$\mathcal{ T}_{\mathrm{loc2}}$. Note that vertical scales in the last two
panels are different from each other and from vertical scales in the first
three panels. The ranges of arguments in these figures are limited due to
various inequalities obeyed by the momenta. In particular,
\begin{equation}
k_{12}^{2}+k_{14}^{2}\leq \sum \limits_{i=1}^{4}k_{i}^{2}\;,
\nonumber
\end{equation}
\begin{equation}
\sqrt{k_{1}^{2}+k_{4}^{2}-2k_{1}k_{4}}\leq k_{14}\leq
\sqrt{k_{1}^{2}+k_{4}^{2}+2k_{1}k_{4}}\;.
\nonumber
\end{equation}

The limits we present are:

1. Equilateral limit,   $k_{1}=k_{2}=k_{3}=k_{4}$. We plot in
 Fig.~\ref{F:eq} the trispectra as functions of $k_{12}/k_1$ and
$k_{14}/k_1$. Clearly seen is the singularity $\mathcal{ T} \propto
 k_{12}^{-1}, k_{14}^{-1}$ in our trispectrum and in its singular part, as
well as stronger singularity $\mathcal{ T}_{\mathrm{loc1}} \propto
 k_{12}^{-3}, k_{14}^{-3}$ in the first local trispectrum. As we pointed
 out in Section~\ref{sec:Intro}, inflationary models produce trispectra
without singularities at $k_{12} \to 0$ and/or $k_{14} \to
0$~\cite{Chen:2009bc,Seery:2006vu,Seery:2008ax,Arroja:2009pd,Bartolo:2010di}.
Thus, the singularity, which is due to the infrared enhancement of the
modes $\delta \rho$, is a distinctive feature of the conformal models.

2. Specialized planar limit, $k_{1}=k_{3}=k_{14}$ and
\[
k_{12}=\left[k_{1}^{2}+\displaystyle\frac{k_{2}k_{4}}{2k_{1}^{2}}\left(k_{2}k_{4}+
\displaystyle\sqrt{(4k_{1}^{2}-k_{2}^{2})(4k_{1}^{2}-k_{4}^{2})} \right)
\right]^{1/2} \; .
\]
The trispectra are shown in Fig.~\ref{F:SPL} as functions of $k_2/k_1$
and $k_4/k_1$. The structures along the diagonal are again due to
the singularity, now at $k_{13} \to 0$, which corresponds to $k_2 \to k_4$.
Note that our total trispectrum vanishes at the boundaries $k_2=0$ and
$k_4=0$ (this can be established analytically). The latter feature is
similar to many inflationary models~\cite{Chen:2009bc,Bartolo:2010di},
while it is absent for $\mathcal{ T}_{\mathrm{loc1}}$ and  $\mathcal{
T}_{\mathrm{loc2}}$.

3. Near the double-squeezed limit, $k_{3}=k_{4}=k_{12}$. We show in
Fig.~\ref{F:NDSL} the combinations
\[
\frac{\mathcal{ T}}{\prod \limits_{i=1}^{4}k_{i}} \; ,
\]
and in Fig.~\ref{F:NDSL-2} the trispectra $\mathcal{ T}$ themselves as
functions of $k_2/k_1$ and $k_4/k_1$. Clearly, our trispectrum is quite
different from local ones in this limit.

\section{Conclusions}
\label{sec:concl}

By comparing the upper panels in Figs.~\ref{F:eq}, \ref{F:SPL},
\ref{F:NDSL}, \ref{F:NDSL-2} one observes that the most notable features
of the trispectrum in conformal models are well captured by its singular
part, which has the factorized form \eqref{may19-1}. It is also clear that
the trispectrum is substantially different from the trispectra of local
forms. Furthermore, the comparison of our trispectrum with the trispectra
given, e.g., in Refs.~\cite{Chen:2009bc,Bartolo:2010di} shows that our
trispectrum is considerably different from the trispectra inherent in
inflationary models. Hence, the shape of the non-Gaussianity, together
with the statistical anisotropy, is an interesting signature of the
conformal mechanisms.

As we observed in Section~\ref{sec:general}, models employing conformal
invariance for generating the flat scalar power spectrum are
indistinguishable at the leading non-linear order. It remains to be
understood whether one can discriminate between concrete models from this
class, even in principle. In any case, it would be extremely interesting
to learn (or rule out) that, in a certain sense, our Universe started out
conformal.

\textit{Note added.} After this work has been published in arXive, the
paper~\cite{Hinterbichler:2011qk} appeared, where the results similar to
those presented in Section~\ref{Section/Pg5/1:paper/Perturbations in
conformal scenarios} were obtained in much more general context. We are
indebted to K.~Hinterbichler and J.~Khoury for making their paper
available to us.

\section*{Acknowledgements} The authors are indebted to S.~Ramazanov for
useful comments and discussions. This work has been supported in part by
the Federal Agency for Science and Innovations under state contract
02.740.11.0244 and by the grant of the President of the Russian Federation
NS-5525.2010.2. The work of M.L. has been supported in part by RFBR grant
11-02-92108. The work of S.M. has been supported in part by RFBR grant
11-02-01220. M.L. and S.M. acknowledge the support by the Dynasty
Foundation. The work of V.R. has been supported in part by the SCOPES
program.

\unitlength=1mm

\unitlength=1mm

\unitlength=1mm

\unitlength=1mm

\newpage \section*{Appendix}

In this Appendix we perform the complete calculation of the tripsectrum.
Notably, this can be done analytically.

\begin{figure}
\begin{center}
\includegraphics[width=70mm]{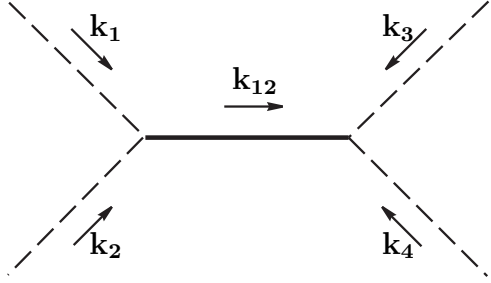}
\end{center}
\caption{
\label{!!}
$s$-channel diagram that contributes to the trispectrum.
Dashed and solid lines correspond to the two-point functions of the fields
$\theta$ and $\delta \rho$, respectively.}
\end{figure}

The computation is conveniently performed in terms of
symmetric polynomials. For the $s$-channel
diagram of Fig.~\ref{!!} these are
\begin{equation}
K_{1}=k_{1}+k_{2}\;,\ \ K_{2}=k_{1}k_{2}\;,\ \ P_{1}=k_{3}+k_{4}\;,\ \
P_{2}=k_{3}k_{4} \; . \nonumber
\end{equation}
\unitlength=1mm
In what follows, we encounter the combinations
\begin{equation}
Z_{K_{1},K_{2}}=K_{1} (3 K_{1}^2 - 12 K_{2} - k_{12}^2)\;,\ \ \ \
X_{K_{1},K_{2}}=k_{12}^{4}+2k_{12}^{2}(K_{1}^{2}-2K_{2})
+12K_{1}^{2}K_{2}-3K_{1}^{4} \;.\nonumber
\end{equation}
In these notations, the expressions \eqref{Eq/Pg8/1:in-in} and
\eqref{Eqn/Pg8/1:in-in} read
\begin{eqnarray}
\tilde{R} &=& [K_{1}^2 - 2 K_{2} - k_{12}^2 - K_{2} (K_{1}^2 - k_{12}^2)
\xi ^2] \cos(K_{1}  \xi ) \nonumber\\
&+&K_{1} (K_{1}^2 - 2 K_{2} - k_{12}^2) \xi  \sin(K_{1}  \xi ) \; ,
\nonumber
\end{eqnarray}
\begin{eqnarray}
\tilde{Q}&=& [K_{1}^2 - 2 K_{2} - k_{12}^2 - K_{2} (K_{1}^2 - k_{12}^2)
\xi ^2] \sin(K_{1}  \xi) \nonumber\\
&-&K_{1} (K_{1}^2 - 2 K_{2} - k_{12}^2) \xi \cos(K_{1}  \xi ) \; .
\nonumber
\end{eqnarray}
The terms \eqref{Eq/Pg5/125:in-in} give the following contribution to the
four-point function:
\begin{equation}
G_{4 \ T \bar{T}} = \frac{h^{2} }{\prod \limits_{i=1}^{4}k_{i}^{3}}
\frac{\pi }{32}\left[ \phantom{\frac{\phantom{1}}{\phantom{1}}} \mathcal{
U}(k_{1},k_{2},k_3,k_4,k_{12}) + 2~\mbox{permutations}
\phantom{\frac{\phantom{1}}{\phantom{1}}}\right] \; , \nonumber
\end{equation}
where, in self-explaining notations,
\begin{equation}
\mathcal{ U}(k_{1},k_{2},k_3,k_4,k_{12}) =\frac{32}{\pi }\int
\limits_{0}^{\infty }d\xi U(k_{1},k_{2},k_3,k_4,k_{12};\xi ) \nonumber
\end{equation}
and
\begin{eqnarray}
U(k_{1},k_{2},k_3,k_4,k_{12};\xi )&=&\int \limits_{0}^{\xi }
\frac{d\chi}{\sqrt{\xi \chi}} \left(J_{5/2}(k_{12}\xi )Y_{5/2}(k_{12}\chi
) \phantom{\frac{\phantom{1}}{\phantom{1}}}\!\!\! -\!\!\!
\phantom{\frac{\phantom{1}}{\phantom{1}}} Y_{5/2}(k_{12}\xi
)J_{5/2}(k_{12}\chi ) \right) \nonumber\\
&&\times\left(\tilde{Q}_{k_{1},k_{2},k_{12}}(\xi )
\tilde{R}_{k_3,k_4,k_{12}}(\chi  )
 \phantom{\frac{\phantom{1}}{\phantom{1}}}\!\!\! +
 \!\!\!\phantom{\frac{\phantom{1}}{\phantom{1}}}
\tilde{Q}_{k_3,k_4,k_{12}}(\xi ) \tilde{R}_{k_{1},k_{2},k_{12}}(\chi  )
\right)\;. \nonumber
\end{eqnarray}
A tedious but straightforward calculation of the latter integral gives
\[
U = U_0 + U_I \; ,
\]
where
\begin{eqnarray}
U_{0}&=& \frac{\tilde{Q}_{k_3,k_4,k_{12}}(\xi )}{2\pi k_{12}^{5}\xi
^{3}}\left(2k_{12}^{3}\xi \cos(K_{1}\xi )(2K_{2}k_{12}^{2}\xi
^{2}-3(K_{1}^{2}-4K_{2}+k_{12}^{2}))
\phantom{\frac{\phantom{1}}{\phantom{1}}} \right.\nonumber\\
&&\left. \phantom{\frac{\phantom{1}}{\phantom{1}}}
-2K_{1}k_{12}\sin(K_{1}\xi )(2k_{12}^{4}\xi
^{2}+3k_{12}^{2}+36K_{2}-9K_{1}^{2})\right)+\{k_{1},k_2\leftrightarrow
k_3,k_4\},
\label{Eqn/Pg18/5:in-in}\\
U_{I}&=& \frac{\tilde{Q}_{k_3,k_4,k_{12}}(\xi )X_{K_{1},K_{2}}}{2\pi
k_{12}^{5}\xi ^{3}}\left(\left[(k_{12}^{2}\xi ^{2}-3)\sin(k_{12}\xi
)+3k_{12}\xi \cos(k_{12}\xi )\right]
\phantom{\frac{\phantom{1}}{\phantom{1}}} \right. \nonumber\\
 &&\times \left. \left[\mathrm{Ci}((K_{1}-k_{12})\xi
)+\mathrm{Ci}((K_{1}+k_{12})\xi )\right] \right.\nonumber
\\
&+&\left.\left[(k_{12}^{2}\xi ^{2}-3)\cos(k_{12}\xi )-3k_{12}\xi
\sin(k_{12}\xi )\right] \right. \nonumber \\
 &&\left. \phantom{\frac{\phantom{1}}{\phantom{1}}} \times
 \left[\mathrm{Si}((K_{1}-k_{12})\xi )-\mathrm{Si}((K_{1}+k_{12})\xi
   )\right] \right)+ \{k_{1},k_2\leftrightarrow k_3,k_4\} \; . \nonumber
 \end{eqnarray}
Note that $U_{0}$ contains only trigonometric functions and powers of
$\xi$. So, the integration of $U_{0}$ over $\xi $  is cumbersome but
straightforward\footnote{At $\xi \to 0$ the original integrals converge.
Nevertheless, some particular terms may produce divergences (note, e.g.,
that $U_{0}$ contains $\cos(K_1 \xi) /\xi $). To regularize these
divergences we integrate over $\xi $ from $\alpha >0$ and take the limit
$\alpha \to 0$ in the end of the calculation.}. On the other hand, $U_I$
involves cosine and sine integrals  Ci and Si, so the integration of
it is tricky. We perform the latter integration by making use of the
integral representations of the functions Si and Ci and changing the order
of integration, i.e., we first integrate over $\xi $ and then integrate
over $\chi $:
 \begin{align}
\mathcal{ U}_{I}&=\frac{32}{\pi }\int \limits_{0}^{\infty }d\xi U_{I}(\xi
)\nonumber\\
&= \frac{32}{\pi ^{2}}\int \limits_{0}^{\infty }\frac{\cos(K_{1}\chi)d\chi
}{\chi } \int \limits_{0}^{\chi }d\xi \frac{\tilde{Q}_{k_3,k_4,k_{12}}(\xi
)X_{K_{1},K_{2}}}{ k_{12}^{5}\xi ^{3}}\nonumber\\
&\times\left(\left[(3-k_{12}^{2}\xi ^{2})\sin(k_{12}\xi )-3k_{12}\xi
\cos(k_{12}\xi )\right]\cos(k_{12}\chi  )
 \phantom{\frac{\phantom{1}}{\phantom{1}}} \right.\nonumber\\
 &\left. \phantom{\frac{\phantom{1}}{\phantom{1}}}
 -\left[(3-k_{12}^{2}\xi ^{2})\cos(k_{12}\xi )+3k_{12}\xi \sin(k_{12}\xi
)\right] \sin(k_{12}\chi ) \right)+ \{k_{1},k_2\leftrightarrow k_3,k_4\}
\nonumber\\
&=\mathcal{ U}_{I0}+\mathcal{ U}_{II}. \nonumber
\end{align}
The inner integration over $\xi $ is lengthy but again straightforward. It
again produces two types of contributions: the first one does not contain
sine and cosine integrals,
\begin{align}
\mathcal{ U}_{I0}&=\frac{16X_{K_{1},K_{2}}}{\pi^{2} k_{12}^{4}}\int
\limits_{0}^{\infty } \frac{\cos(K_{1}\chi)d\chi  }{\chi ^{2}}\left( 2
P_{1} (k_{12}^{2}-3 P_{2}) \chi \cos(P_{1} \chi) + P_{1}
(k_{12}^{2}+12P_{2}-3 P_{1}^2 ) \chi \cos( k_{12} \chi)
\phantom{\frac{\phantom{1}}{\phantom{1}}} \right.\nonumber\\
&\left. \phantom{\frac{\phantom{1}}{\phantom{1}}} + [3 P_{1}^2 - 6
 P_{2} - k_{12}^{2}(3  -2 P_{2} \chi^2) ]\sin(P_{1} \chi) \right)
+\{k_{1},k_2\leftrightarrow k_3,k_4\} \; ,
\label{Eqn/Pg20/1:in-in}
\end{align}
while the second one contains Si and Ci in the integrand,
\begin{align}
&\mathcal{ U}_{II}=\frac{8X_{K_{1},K_{2}}X_{P_{1},P_{2}}}{\pi^{2}
k_{12}^{5}}\int \limits_{0}^{\infty } \frac{\cos(K_{1}\chi )d\chi }{\chi
}\left(\cos(k_{12}\chi )\left[\mathrm{Ci}((P_{1}+k_{12})\chi
)-\mathrm{Ci}((P_{1}-k_{12})\chi )
\phantom{\frac{P1}{P1}}\right.\right.\nonumber\\
&-\left.\left. \log\left( \frac{P_{1}+k_{12}}{P_{1}-k_{12}}\right)
\right]+\sin(k_{12}\chi )\left[ \mathrm{Si}((P_{1}+k_{12})\chi
)+\mathrm{Si}((P_{1}-k_{12})\chi )  \right] \right)
+\{k_{1},k_2\leftrightarrow k_3,k_4\} \; .
\label{Eqn/Pg20/2:in-in}
\end{align}
Nevertheless, both of these integrals can be evaluated analytically, the
relevant formulas being
\begin{eqnarray}
S_{i}(a,b,c)&=&\lim_{\epsilon \to +0}\int \limits_{0}^{\infty
}dx\mathrm{e}^{-\epsilon
x}\frac{\sin(ax)\mathrm{Si}(bx)}{x}+\{b\rightarrow c\} \nonumber\\
&=&\frac{1}{4}\left[\Theta (b-a)L_{2}^{*}(b/a)+\Theta
(a-b)L_{2}(b/a)-2L_{2}(-b/a)+L_{2}(b/a) \right]+\{b\rightarrow c\}\;,
\nonumber
\end{eqnarray}
\begin{align}
C_{i}(a,b,c)&=\lim_{\epsilon \to +0}\int \limits_{0}^{\infty
}dx\mathrm{e}^{-\epsilon
x}\frac{\cos(ax)[\mathrm{Ci}(bx)-\mathrm{Ci}(cx)-\log(b/c)]}{x}
\nonumber\\
&=\frac{1}{8}\left\{\log\left(\frac{b^{2}}{c^{2}} \right)\log\left(
\frac{(ia)^{2}}{bc}\right)
+L_{2}\left(\frac{b^{2}}{a^{2}}\right)-L_{2}\left(\frac{a^{2}}{b^{2}}\right)-
L_{2}\left(\frac{c^{2}}{a^{2}}\right)+
L_{2}\left(\frac{a^{2}}{c^{2}}\right) \right\}\;. \nonumber
\end{align}
where $L_{2}$ is the dilogarithm function. In this way we obtain
\begin{align}
\mathcal{ U}&=\frac{1}{8\pi
k_{12}^{4}}\left(Z_{k_{1},k_{2}}X_{k_3,k_4}\log
\left[\frac{(P_{1}^{2}-k_{12}^{2})^{2}}{(K_{1}+P_{1})^{4}} \right] +
Z_{k_3,k_4}X_{k_{1},k_{2}}\log
\left[\frac{(K_{1}^{2}-k_{12}^{2})^{2}}{(K_{1}+P_{1})^{4}} \right]
\right.\nonumber\\
&+\left. \frac{k_{12}^{2}}{(K_{1}+P_{1})^{3}}[{\cal G}_{2}+k_{12}^{2}
{\cal G}_{4}+k_{12}^{4}{\cal G}_{6}] \right) \nonumber\\
&+\frac{1}{8\pi  k_{12}^{5}}X_{k_{1},k_{2}}X_{k_3,k_4}
\left[ C_{i}(K_{1}+k_{12},P_{1}+k_{12},P_{1}-k_{12}) +
C_{i}(K_{1}-k_{12},P_{1}+k_{12},P_{1}-k_{12})
\phantom{\frac{\phantom{1}}{\phantom{1}}} \right.\nonumber\\
&+\left. S_{i}(K_{1}+k_{12},P_{1}+k_{12},P_{1}-k_{12})
-S_{i}(K_{1}-k_{12},P_{1}+k_{12},P_{1}-k_{12})+ \{k_{1},k_2\leftrightarrow
k_3,k_4\} \!\!\!\phantom{\frac{\phantom{1}}{\phantom{1}}} \right]\;.
\nonumber
\end{align}
The first two lines here come from the integration of $U_0$,
Eq.~\eqref{Eqn/Pg18/5:in-in}, and from (\ref{Eqn/Pg20/1:in-in}), while the
rest is due to \eqref{Eqn/Pg20/2:in-in}. The notations are
\begin{eqnarray}
{\cal G}_{2}(k_{1},k_{2},k_3,k_4)&=&
-12(K_{1}+P_{1})^{4}(K_{1}^{2}-4K_{2})(P_{1}^{2}-4P_{2}) \;,\nonumber
\\
{\cal G}_{4}(k_{1},k_{2},k_3,k_4)&=&
32K_{2}P_{2}(K_{1}^{2}+K_{1}P_{1}+P_{1}^{2}) \nonumber\\
& -&16P_{2} (K_{1}+P_{1})(K_{1}^3 + 2 K_{1}^2 P_{1} - 3 K_{1} P_{1}^2 -
3 P_{1}^3)
\nonumber\\
&-&16K_{2} (K_{1}+P_{1})(P_{1}^3 + 2 P_{1}^2 K_{1} - 3 P_{1} K_{1}^2 -
3 K_{1}^3)
\nonumber\\
& -&4(K_{1}+P_{1})^{2}(3K_{1}^{4}-2K_{1}^{2}P_{1}^{2}+3P_{1}^{4})
\nonumber
\;,\\
{\cal G}_{6}(k_{1},k_{2},k_3,k_4)&=&-32K_{2}P_{2}-16(K_{1}+P_{1})\left[
K_{2}(K_{1}+2P_{1}) +P_{2}(P_{1}+2K_{1}) \right]\nonumber\\
&+&4(K_{1}+P_{1})^{2}(3K_{1}^{2}+2K_{1}P_{1}+3P_{1}^{2}) \;.\nonumber
\end{eqnarray}
We add the part of \eqref{Eq/Pg5/1:in-in} that corresponds to the
 $s$-channel diagram of Fig.~\ref{!!} and is given by
\eqref{may26-10}, and obtain
the overall contribution of the $s$-channel diagram:
\begin{eqnarray}
S(k_{1},k_{2},k_{3},k_{4},k_{12})&=&\frac{h^{2}}{\prod
\limits_{i=1}^{4}k_{i}^{3}} \frac{\pi }{32} \left[ \mathcal{
Y}(k_{1},k_{2},k_{12}) \mathcal{ Y}(k_{3},k_{4},k_{12}) +
\mathcal{ J}(k_{1},k_{2},k_{12}) \mathcal{ J}(k_{3},k_{4},k_{12})
\phantom{\frac{\phantom{1}}{\phantom{1}}} \right. \nonumber\\
&& \left.\phantom{\frac{\phantom{1}}{\phantom{1}}} +\mathcal{
U}(k_{1},k_{2},k_{3},k_{4},k_{12}) \right] \;.
\label{Eq/Pg15/2:in-in}
\end{eqnarray}
Together with crossing terms
this yields our final result
\begin{eqnarray}
G_{4}(\mathbf{k}_{1},\mathbf{k}_{2},\mathbf{k}_{3},\mathbf{k}_{4}) &=&
\frac{h^{2}}{\prod \limits_{i=1}^{4}k_{i}^{3}}\mathcal{
T}(k_{1},k_{2},k_{3},k_{4},k_{12},k_{14})=
S(k_{1},k_{2},k_{3},k_{4},k_{12}) \nonumber\\
&&+ S(k_{1},k_{3},k_{2},k_{4},k_{13})
+S(k_{1},k_{4},k_{3},k_{2},k_{14}).
\label{Eq/Pg16/1:in-in}
\end{eqnarray}

\end{document}